\documentclass[preprint,12pt]{elsarticle}

\newcommand{\KG}{\textcolor{black}}
\providecommand{\KH}[1]{\textcolor{black}{#1}} %
\providecommand{\NP}[1]{\textcolor{black}{#1}} %

\usepackage[utf8]{inputenc}
\usepackage{booktabs} %
\usepackage{multirow}
\usepackage{nicefrac}
\usepackage[caption=false]{subfig} %
\usepackage[font={small}]{caption}
\usepackage{xcolor}
\usepackage{natbib}
\usepackage{url}
\setcitestyle{authoryear,open={(},close={)},aysep={}}

\makeatletter\renewcommand\paragraph{\@startsection{paragraph}{4}{\z@}
  {.5em \@plus1ex \@minus.2ex}{-.5em}{\normalfont\normalsize\bfseries}}\makeatother

\AtBeginDocument{%
  \providecommand\BibTeX{{%
    \normalfont B\kern-0.5em{\scshape i\kern-0.25em b}\kern-0.8em\TeX}}}

\journal{Tourism Management}

\begin{document}

\begin{frontmatter}

\title{Learning Patterns of Tourist Movement and Photography from Geotagged Photos at Archaeological Heritage Sites in Cuzco, Peru}
\author{Nicole D. Payntar$^{1}$ \  Wei-Lin Hsiao$^{1,2}$ \  R. Alan Covey$^{1}$ \  Kristen Grauman$^{1,2}$ 
\vspace{.5em} \\
$^1$The University of Texas at Austin \quad $^2$ Facebook AI Research
\vspace{-.5em}
}

\begin{abstract}
The popularity of media sharing platforms in recent decades has provided an abundance of open source data that remains underutilized %
by heritage scholars. By pairing geotagged internet photographs with machine learning and computer vision algorithms, we build upon the current theoretical discourse of anthropology associated with \emph{visuality} and \emph{heritage tourism} to identify travel patterns across a known archaeological heritage circuit, and quantify visual culture and experiences in Cuzco, Peru. Leveraging large-scale in-the-wild tourist photos, our goals are to (1) understand how the intensification of tourism intersects with heritage regulations and social media, aiding in the articulation of travel patterns across Cuzco's heritage landscape; and to (2) assess how aesthetic preferences and visuality become entangled with the rapidly evolving expectations of tourists, whose travel narratives are curated on social media and grounded in historic site representations.
\end{abstract}

\begin{keyword}
geotagged photos \sep anthropology and computer vision \sep clustering \sep visuality \sep heritage tourism \sep visual recognition \sep Peru

\end{keyword}

\maketitle

\section{Introduction}\label{sec:intro}
The rise of social media in recent decades has led to the large-scale influx of publicly distributed images. Photo-sharing websites like Flickr (est. in 2004) now house over 6 billion images generated by more than 40 million unique users \citep{crandallsnavely}. Originally developed to aid in the organization of photos and enable sharing between users, media-sharing websites offer new and underutilized areas of study for cultural heritage researchers when paired with computer vision and machine learning algorithms. Heritage studies focused on visuality and tourism currently remain theoretically driven, %
with few moving beyond the conceptual stage to incorporate methods that generate empirical evidence and quantify visual culture (i.e. photographs).

Using Cuzco, Peru, as a case study, we present an innovative application of computer vision and machine learning methods to understand archaeological heritage circuits, the evolution of their aesthetic legacies, tourist movement trends (and associated economic influences), and real-world %
visual experiences at heritage sites. To our knowledge, this is the first time that vision and machine learning techniques have been used to study the visual experiences of archaeological heritage tourism. Our goals are two-fold: to understand how the intensification of tourism intersects with heritage regulations and social media, aiding in the articulation of travel patterns across a known heritage landscape; and to assess how aesthetic preferences and visuality become entangled with the rapidly evolving expectations of tourists, whose travel narratives are curated on social media and grounded in historic site representations.

Located in the southwestern Andes mountains of Peru, Cuzco is the former capital of the Inca Empire (1438-1532 CE) and one of the world's best-known tourist regions. Tourism contributes \$7.6 billion annually to the Peruvian economy and provides 3.9\% of Peru's GDP \citep{rice2018making}. Much of Peru's modern tourist economy centers on the Cuzco region and the idea of Andean ``timelessness'' which links the Inca past to the present \citep{covey2017legacies}. The perception of timelessness depicted in Cuzco's imagery aids in the creation of heritage hierarchies, aestheticizes places, and assigns economic and cultural values to heritage landscapes by selecting what is meant to be ``seen'' and preserved \citep{watson2010waterton}. In particular, imagery produced by 19th and 20th century explorers helped to popularize Cuzco as a tourist destination and critically influenced how archaeological sites were curated for mass consumption. The Inca sites first reported by the explorer Hiram Bingham have remained at the core of Cuzco's tourism industry in part due to the sensationalism surrounding Bingham's ``discovery'' of Machu Picchu in 1911. 

Today, archaeological heritage tourism is structured by the \emph{Boleto Turístico del Cuzco} (BTC), a multi-site pass that grants access to 10 sites in and around Cuzco. This includes: Sacsayhuaman, Tipón, Pikillacta, Tambomachay, Puca Pucara, Chinchero, Ollantaytambo, Pisac, Moray, and Qenqo. With the exception of the Wari (600-1000 CE) site of Pikillacta, all of the archaeological properties on the BTC are monumental Inca sites.  Those 10 sites plus %
two others---the UNESCO World Heritage sites of Cuzco and  Machu Picchu---comprise the 12 archaeological sites (Fig.~\ref{fig:cuzco_map}) in the circuit we study.

\begin{figure*}[t]
    \begin{center}
   \includegraphics[width=.95\linewidth]{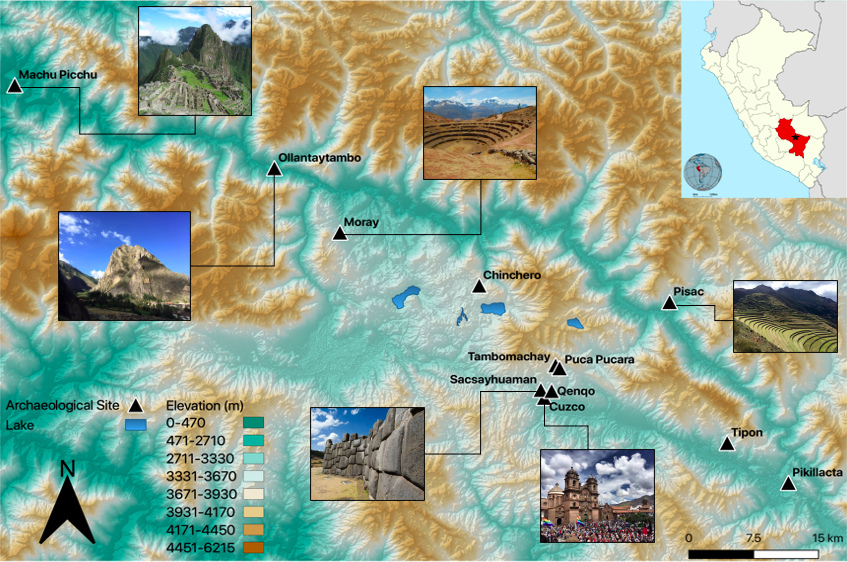}
   \vspace*{-0.05in}
    \caption{Multiscalar map of Cuzco, Peru with photographs representing the six most visited archaeological sites in our study.}
   \label{fig:cuzco_map}
   \end{center}
\end{figure*}

\NP{Our work is interdisciplinary in nature, bridging the fields of Anthropology and Artificial Intelligence (AI) to build on the theoretical underpinnings currently driving studies of visual culture and heritage tourism. We approach the visual experiences, perceptions, and expectations of tourists traveling to the Cuzco region through a study of visuality.} ``Visuality" is a concept in anthropology that incorporates how vision is culturally and historically constructed and is defined by ``how we see, how we are able, allowed, or made to see, and how we see this seeing'' \citep{rosevisuality}. \NP{Heritage tourism is highly visual in nature, with archaeological sites being marketed and packaged to tourists as part of a national ``brand" that is produced and reinforced over time. The images that accompany heritage tourism and those that are shared on social media influence the commodification of landscapes, ruins, and native bodies and constrain what consumers see during short-term visits. They then in turn impact how tourists record, disseminate, and shape heritage images.} For this reason, visuality is a significant framework from which to critically assess the historic trajectory and conceptualization of archaeological heritage circuits generated through imagery.
 
We aim to quantify visuality and tourism patterns across Cuzco's heritage circuit.  To this end, we create a tourist photo dataset \NP{using a community photo collection}, with photos from 2,261
users at 12 sites with associated geotags and timestamps spanning 2004-2019. 
We introduce a series of analyses using computer vision and machine learning to recover insights relevant to anthropology from this imagery.  In particular, 
tourist movement across sites is fit to a Markov model, which
is then applied %
to determine travel patterns between BTC sites via multiple mini-itineraries (day trips), as well as general sequences of movement between BTC and UNESCO sites. The density of unique users, total photos taken, and average time spent by users per site are analyzed to assess \NP{relative popularity among Cuzco's archaeological sites, as well as patterns of movement \emph{between} sites that reflect how multiple, rural heritage sites are bundled into scripted tour itineraries.} Through our analyses, we compare the significance of BTC circuits and individual archaeological heritage sites on the landscape. Finally, to analyze the iconicity of modern tourist photos, we identify major photography themes at each site by clustering in the visual feature space of the photos, and we consider how canonical images define Cuzco's tourist experience and continue to reinforce the region’s %
aesthetic legacy.

We first review related work in Sec.~\ref{sec:related}. Sec.~\ref{sec:hist_arc} provides an overview of archaeological heritage tourism in Cuzco, Peru. Sec.~\ref{sec:approach} introduces our methodology for the case study. In Sec.~\ref{sec:exp}, we present the statistical results and mined patterns of tourist movement and canonical view themes from our \NP{internet photo} dataset.  %
We conclude with a summary of our findings.
\section{Related Work}
\label{sec:related}
Archaeological applications of machine learning have been limited to typological and stylistic analyses of artifacts \citep{horrML, gansellML}, with the majority of studies concerning the representation of peoples and places centered around theoretical discussions of visuality \citep{garrod, burns, urrytouirstgaze, sandvisuality}. 

Scholars in the Cuzco region have made significant theoretical contributions to studies of heritage and visuality \citep{Poolevisual, scorer2014chambi, cuzcotopia}. %
This scholarship has focused on tracing the history of the visual economy and the emergence of new technologies (e.g., mass printing and photography) that transformed the production, circulation, and interpretation of visual culture over the last three centuries \citep{Poolevisual, scorer2014chambi}. Investigations yielded the development of a visual aesthetic for Andean ruins (and indigenous people), whose perceived ``authenticity'' became part of their value within national heritage programs and the tourist economy. This paper builds on these thematic narratives, applying machine learning methods to the current standard for visual dissemination---internet-based digital images---to generate empirical patterns that can link current practices to the overall historical arc of exploration and heritage tourism in Cuzco.

More recently, heritage studies have taken a top-down approach to visuality using aerial photography, satellite imagery, and drones. These remote sensing techniques have become popular archaeological tools for ``seeing'', monitoring, and discovering archaeological heritage sites from afar \citep{casana, bevanlake, parcak}. Remote sensing imagery tends to be collected at larger scales by governments, private companies, and scientific researchers and it is typically designed to capture landscape and site characteristics, not to see how tourists use those places. More relevant to our work, publicly posted images, especially those shared on social \NP{media platforms and community photo websites}, offer a distinct visual resource.
These photos are accompanied by geotags, timestamps, and user IDs,
which permit the direct observation of how tourists move within and between heritage sites, study tourist movement in national parks and urban cities), as well as which images they choose to capture and post online \citep{spatiotemporalhotspot2018,zheng2012mining,touristsocail2019}. 

In the computer vision literature, the popularity of %
photo sharing platforms has opened up new possibilities to understand our world at a lower cost and %
in a more automated way.
For example, at popular tourist sites like Trevi Fountain and Notre Dame, photos are taken densely from all viewpoints and locations around the sites, %
by large volumes of tourists. Pioneering work %
harnessed this open source data %
to apply structure from motion~\citep{hartley2003multiple} and reconstruct 3D point clouds for entire sites~\citep{snavely2006tourism3D}. 
Multi-modal analysis of large collections offers new ways to visualize tourist trends~\citep{kleinberg}.  
As many tourist photos are street scenes or self-portraits irrelevant to the sites themselves, automatic detection of \emph{iconic images} can help capture key parts of a site,
with applications in photo summarization and browsing~\citep{zhang2018detecting} and landmark segmentation~\citep{simon2008wisdom}.  %
When geotags are not available, %
vision methods can estimate geolocation (latitude and longitude) from the images themselves%
~\citep{hays2008im2gps,Hertzmann2009im2gps,Chen2011CluesFT}.
Timestamps and the photo owner's ID provide temporal context that allows for travel sequence modeling for accurate geolocation~\citep{Hertzmann2009im2gps,Chen2011CluesFT,kleinberg}.

Pushing forward to combine these spatial and temporal findings to gain socio-cultural insights, recent studies have measured and tracked the occurrence of ecological phenomena \citep{zhang2012mining,wang2013observing,wang2016tracking}, while also successfully %
estimating the demography of $200$ US cities by using $50$ million images of street scenes gathered with Google Street View cars \citep{gebru2015bigvision,gebru2017nas}. Prior to such applications, these results were only available through manual processing %
which does not scale well or update in real-time. \NP{Recently, scholars have turned toward social media and visual content analysis using geotagged photos to understand tourist behavior and perceptions at destination hotspots \citep{Zhang2019, Vu2015, Stepchenkova2013, Kaufman2019}}.
At a high level, our work is in a related socio-cultural vein.  However, we are the first to explore machine learning for heritage tourism studies, %
specifically for heritage circuits and visual culture in Cuzco, Peru.
Our findings of tourist flows within a known circuit provide insights to heritage conservation, while our content analysis of tourist photos sheds light on how modern peoples visually experience Cuzco's heritage landscape and reproduce the region's aesthetic legacy.

\section{Historic Arc of Inca Archaeological Tourism in Cuzco}
\label{sec:hist_arc}

Next we present historical context for our inter-disciplinary study. 
Following more than a century of foreign expeditions to the Cuzco region, archaeological tourism to the former Inca heartland took root in the early 1900s \citep{rice2018making}. Hiram Bingham represents the nexus of the expeditionary aspiration to ``discover'' something unknown and introduce it to the developed world and the more scripted itinerary of the tourist. Through his relationship with \emph{National Geographic Magazine} \citep{bingham1913wonderland}, Bingham shared his early encounters with the “lost city” of Machu Picchu, but his expeditions also helped to crystallize the modern touristic route in Cuzco by illustrating and describing key sites, including Sacsayhuaman, Tambomachay, Puca Pucara, Ollantaytambo, Pisac, Tipón, and Pikillacta \citep{bingham1922highlands}. The vivid imagery of Bingham's publications encouraged growing numbers of foreigners to visit the Cuzco region as tourists, with guidebooks appearing by the early 1920s \citep{bauer2004heartland}. %
Over the next decade, government officials recognized the economic potential of tourism and began to fund reconstruction projects at tourist sites, including Sacsayhuaman, Tambomachay, Pikillacta, and Pisac.  %

Following a powerful earthquake that destroyed most of Cuzco's modern buildings in 1950, archaeological tourism reemerged %
with the help of the newly formed United Nations Educational, Scientific, and Cultural Organization (UNESCO).
Annual tourist visits to the Cuzco region topped 100,000 by the mid-1970s \citep{steel2008vulnerable} and local archaeologists and officials sought to develop broader itineraries that would encourage tourists to spend more time (and money) in the Inca heartland. Established in 1978, Cuzco's original Boleto Turístico (BTC) allowed 10-day visitor access to the region’s top archaeological attractions \citep{rice2018making}. Tourist initiatives in the 1970's also included the promotion of the Inca Trail, a heritage route to Machu Picchu \citep{maxwell2012tourism}, %
which along with the monumental center of Cuzco became the first Peruvian properties to be added to UNESCO’s World Heritage List in 1983. %

Today, roughly $90\%$ of all international visits to Peru feature a stop in Cuzco, with half of these including trips to Machu Picchu \citep{larsonpoudyal}. Machu Picchu %
\KG{is} one of the “New7Wonders” and annual visits to the Cuzco region now top 3 million, an increase of more than $500\%$ since 2000 \citep{Dircetur2017}. The influx of tourists to these two UNESCO sites has had the added effect of bolstering sales of the BTC while entrenching and perpetuating Cuzco's established heritage narrative of Inca monumentality.

In 2008, the BTC circuit was expanded to include: a 1-day pass to Sacsayhuaman, Qenqo, Tambomachay, and Puca Pucara (BTC I); a 2-day pass to Cuzco's six museums, and the archaeological sites of Tipón and Pikillacta (BTC II); and a 2-day pass to Pisac, Ollantaytambo, Chinchero, and Moray (BTC III). The addition of the 1- and 2-day passes has contributed to an uptick in visits to Cuzco's more remote sites (e.g., Moray \citep{Dircetur2017}), %
and has also affected movement patterns between BTC sites, \KG{as we will examine quantitatively through image data below.} %

\section{Methodology}
\label{sec:approach}

\KG{Having provided the historical context for our study, we now present our technical methodology to elicit a quantitative story for this heritage circuit, and to connect the last 15 years of tourism to visuality established by early explorers.}  %

We collected data from \NP{community internet photos} %
to obtain images captured at the $12$ archaeological sites in our study (Sacsayhuaman, Tipón, Pikillacta, Tambomachay, Puca Pucara, Chinchero, Ollantaytambo, Pisac, Moray, Qenqo, Machu Picchu, Cuzco) and to create user albums for each unique user ID. %
To detect patterns of tourist movement and analyze their change over time, 
we model tourists' transition sequences with a Markov model~\citep{markov1960}. 
To identify scenes and objects photographed by users at each site and detect patterns of iconicity, a convolutional neural network is used to extract semantic features from images. %
A clustering algorithm is then used to discover common image themes at each site, such as mountains, stonework, alpaca, terracing, etc. \KG{We detail each of these components in the following section.}

\paragraph{Dataset Creation}
Images for all BTC and UNESCO sites were collected \NP{from an internet photo community} for a 15-year period (2004 - 2019). Google Earth was used to establish a central point of latitude and longitude for each site, as well as an estimated spatial extent (km) and buffer.

We downloaded the data in two stages. In \KG{Stage} 1, we collected photos taken at all $12$ archaeological sites \KG{(listed in Sec.~\ref{sec:intro})} by querying with the site's GPS coordinates and taking the top $4,000$ retrieved images
from each site. Images with GPS coordinates beyond the estimated buffer from the central point of the site were later eliminated.
In \KG{Stage} 2, we then expanded this collection to form an album for each user by querying with all unique user IDs from \KG{Stage} 1, and specified to retrieve photos taken within the time the user visited the sites. Each photo includes meta-data like the image's url, owner's ID, geotag, and timestamp. In total, $57,804$ images were collected from $2,261$ users.  %
Table \ref{tab:photos_per_year} shows the total number of photos per site and the number of unique visitors who traveled to the 12 sites. These metrics were used to infer a site's popularity ranking within our dataset. Machu Picchu was found to be the most popular site visited (as it is advertised as the ultimate destination in the Cuzco region), while the Wari site of Pikillacta (the only non-Inca site included on the BTC) was found to be the least popular.

\paragraph{Mining Tourist Traveling Patterns} %
We %
\KG{model} tourists' traveling patterns with a Markov chain, a stochastic model for a sequence of events that assumes that the probability of each event depends only on the state attained in the previous event \citep{Geron2017}. In our study, an event is a site visit (e.g., Cuzco), and a sequence is the order of sites a tourist traveled (e.g., start in Cuzco and go to Sacsayhuaman and finally to Machu Picchu). 
With photos associated with each unique user (tourist) and site, we  represent the tourist traveling pattern among all sites with a Markov matrix ${P}$. 
It is a square matrix that describes the transitions of a Markov chain, where 
\KG{the ij-th entry} 
${P}_{ij}$ is the probability of moving from site $i$ to site $j$. Each row thus sums to $1$.
We calculate the transition frequencies from site to site \KG{to obtain these} probabilities.
This Markov matrix 
allows us to discover popular transitions between sites, %
how transitions are affected by %
policies regulating heritage landscapes (e.g., the release of the BTC ticket package (Sec. \ref{sec:exp})), and how transitions change over time, as we will discuss below.

\paragraph{Discovering Canonical View Themes}
\label{sec:view_analysis}
Deep neural networks trained on large scale data \KG{provide} a powerful and compact representation. 
\KG{In vision,} features extracted from deep convolutional neural networks (CNN) are widely used to capture objects and scenes in an image \citep{decaf2014, cnnfeat2014, rcnn2014}. Here, we adopt a \KG{successful} CNN called ResNet50~\citep{resnet}, which is pre-trained on ImageNet \citep{imagenet}, a database with more than $10$ million images labeled with $1$K common object categories. \KG{We} extract our image features using its penultimate \KG{layer,} %
since higher layers are known to capture semantic features such as object categories. Pre-training the visual encoder enriches the features to capture common low -- and high -- level visual patterns, beyond mere pixels. \NP{By utilizing a CNN pre-trained on ImageNet for our t-sne and clustering analysis, we are able to identify multiple real-world objects simultaneously without deploying different algorithms for each object and without the added cost of manual labor (for which the task would be impossible).} %

With this rich representation, our goal is to find common themes of canonical views taken at each archaeological site. 
\emph{Clustering} is a data mining technique that groups objects in a way such that those in the same group (called a cluster) are more similar (in some sense) to each other than to those in other groups (clusters).
As such, each cluster corresponds to a discovered theme, and all photos taken at a site can be captured by the representative images, i.e.~\emph{exemplars}, from all themes.

To this end, 
we adopt affinity propagation (AP)~\citep{APcluster2007} clustering.  AP views data points as nodes in a network and exchanges messages between nodes until a set of good quality exemplars emerges. It has the benefits of automatically deciding the number of clusters in a data set, and also representing clusters as \emph{exemplars} instead of a mean/median of the cluster.
\KG{Fig.~\ref{fig:full_clusters} shows} examples of \KG{our} discovered themes.

Aside from discovering canonical views with clustering, we also explore the statistics of which types of scenes are observed across the photos at each site. Specifically, to compute our scene-site occurance matrix (which we will present below in Fig.~\ref{fig:scene_labels}) we use ResNet50 pre-trained on MIT Places \citep{zhou2017places}.
\NP{MIT Places consists of over $10$ million images and $400$ unique scene categories. Because ImageNet is an object-centric database, it is not possible to extract scene labels, making it necessary to also train on MIT Places for this specific task. While some of the general AI classifications associated with the Places database reflect non-Andean categories (e.g.\ medina, rice paddy, catacomb, etc.), shared features remain high and do correlate to Andean life-ways (e.g.\ market, agriculture, tombs, etc.).} %

\section{Results and Analysis}
\label{sec:exp}

We  analyze \KG{(1)} tourists' statistics at each site, %
\KG{(2)} whether tourist transition patterns 
\KG{may be attributable to} available BTC tickets, and \KG{(3) %
the discovered canonical view themes.}

\NP{Table \ref{tab:photos_per_year} shows the total number of photos taken per site per year and Table \ref{tab:site_pop_table} shows site popularity based on the total number of photos taken per site and the total number of unique visitors.} While our best efforts were made to avoid sampling bias and statistical oversight, not all tourists traveling to BTC and UNESCO sites take the same amount of photos, geotag photos, or even upload photos to \NP{community photo collections}. %
For these reasons, biases may exist when calculating the average time a user spent at each site since our calculations are dependent on photo timestamps. This may also create biases when generating site popularity rankings across Cuzco's heritage landscape. \NP{Despite these factors, our quantitative historical study is valuable for addressing tourist practices at a key moment of economic (BTC elaboration and mass tourism), social (share-ability of personal imagery), and technological (ubiquity of smart phones with cameras) change in the Cuzco region.}

\begin{table*} %
  \centering
  \captionsetup[subfigure]{labelformat=empty}
  \footnotesize
  \caption{Total number of photos taken per site from 2004-2019. The most popular photo years for each site are shown in bold.} 
  \subfloat[] {
  \begin{tabular}{c|p{8mm}ccccc@{}}
    Year & \parbox{8mm}{Machu\\Picchu} & Cuzco   &  Sacsayhuaman & Ollantaytambo & Pisac & Chinchero\\
    \midrule
    2019 & 293        &  71         & 38   & 6     & 4    & 0      \\     
    2018 & 1768       &  675        & 96   &  232  & 172  & 57     \\    
    2017 & 2697       &  789        & 326  & 369   & 410  & \bf{205}  \\    
    2016 & 2365       &  719        & 330  & 409   & 224  & 149     \\    
    2015 & 2903       &  1296       & 434  & 268   & 160  & 79      \\    
    2014 & 2601       &  1245       & 542  & 229   & 308  & 74      \\    
    2013 & 3376       &  1157       & \bf{609}  & \bf{515}   & \bf{530}  & 150   \\
    2012 & \bf{4050}  &  \bf{1852}  & 362  & 439   & 399  & 50      \\
    2011 & 2649       &  1005       & 186  & 164   & 132  & 34      \\
    2010 & 1899       &  756        & 221  & 230   & 201  & 87      \\
    2009 & 1678       &  775        & 225  & 123   & 123  & 43      \\
    2008 & 2794       &  848        & 327  & 174   & 292  & 13      \\
    2007 & 1073       &  380        & 188  & 78    & 42   & 29      \\
    2006 & 619        &  131        & 53   & 21    & 57   & 14      \\
    2005 & 651        &  124        & 8    & 28    & 15   & 0       \\
    2004 & 134        &  19         & 13   & 3     & 8    & 0       \\
  \end{tabular}
  }\hfill
  \subfloat[] {
  \begin{tabular}{c|ccccccc@{}}
    Year &  Moray    & Qenqo  & Puca Pucara & Tambomachay & Tipón & Pikillacta & \bf{Total}  \\
    \midrule
    2019  &  0         & 0         & 13        & 0       & 0    & 0   & 425  \\     
    2018  & 75        & 9         & 21        & 15      & 0    & 1   & 3121  \\    
    2017  & 122       & 62        & 72        & \bf{80} & 61   & 12  & 5205  \\    
    2016  & 73        & 31        & 33        & 36      & 46   & 11  & 4426  \\    
    2015  & 76        & 40        & 42        & 33      & 46   & \bf{100} & 5477  \\    
    2014  & 50        & 57        & 34        & 42      & 7    & 7   & 5196  \\    
    2013  & \bf{161}  & 84        & \bf{116}  & 55      & 54   & 43  & 6850  \\
    2012  & 35        & 48        & 59        & 30      & 13   & 1   & 7338  \\
    2011  & 64        & 31        & 49        & 39      & 77   & 38  & 4468  \\
    2010  & 67        & 13        & 19        & 12      & 8    & 18  & 3531  \\
    2009  & 35        & 47        & 48        & 51      & 28   & 0   & 3176  \\
    2008  & 36        & \bf{108}  & 23        & 54      & 1    & 3   & 4673  \\
    2007  & 11        & 17        & 10        & 1       & \bf{82}   & 50   & 1965  \\
    2006  & 25        & 7         & 0         & 0       & 5    & 4   & 936  \\
    2005  & 10        & 1         & 1         & 1       & 0    & 0   & 839  \\
    2004  & 0         & 0         & 0         & 1       & 0    & 0   & 178  \\
  \end{tabular}
  }
   \label{tab:photos_per_year}
\end{table*}

\begin{table*} %
  \centering
  \captionsetup[subfigure]{labelformat=empty}
  \footnotesize
  \caption{Total number of photos taken per site and total number of unique visitors per site. Site popularity (ranked) is inferred from these numbers.}
  \setlength\tabcolsep{4pt}
  \begin{tabular}{c|ccc@{}}
    Site & $\#$ photos & $\#$ unique visitors   &  popularity rank  \\
    \midrule
    Machu Picchu & 31550 & 1498 & 1 \\
    Cuzco & 11842 & 1273 & 2 \\
    Sacsayhuaman & 3958 & 524 & 3\\
    Ollantaytambo & 3288 & 381 & 4\\
    Pisac & 3077 & 348 & 5 \\
    Moray & 840 & 156 & 6 \\
    Chinchero & 984 & 143 & 7 \\
    Qenqo & 555 & 139 & 8 \\
    Puca Pucara & 540 & 125 & 9 \\
    Tambomachay & 454 & 108 & 10 \\
    Tipón & 428 & 34 & 11 \\
    Pikillacata & 288 & 27 & 12 \\
  \end{tabular}
  \label{tab:site_pop_table}
\end{table*}

\NP{Community photo collections} have risen in popularity since 2004 and have led to an increase in data (moving toward the present) as more users sign up to organize, store, and share their images. As seen in Table \ref{tab:photos_per_year}, the number of photos per year ranged significantly across sites over time. The least amount of data collected for our study corresponded to the earliest collection year (2004), while the year with the greatest amount of data available \emph{across} sites was 2013. The largest number of \emph{total} photos per year in our dataset can be found in 2012. There is also the question of whether users choose to make their photos publicly available, further influencing sampling strategies and the overall results of our analysis. Although it is not a perfect vehicle to approach how \emph{all} tourists move across Cuzco's heritage sites, our internet photo archive reflects a dynamic reality of publicly available images.

\paragraph{Popularity of Sites}
Site popularity was determined by comparing the total number of images gathered per site versus the total number of unique visitors per site. %
Moreover, the average amount of time a user stayed at a site was calculated from user albums %
collected in \KG{Stage} 2. %
The earliest and latest time a photo 
was taken at each site were identified for each user. 
Time elapsed from the earliest to the latest time %
\KG{serves} as the total amount of time a user spent at a given site.%

A density map was generated \NP{(using ArcGIS software)} from the total number of photos taken by users at \KG{all 12 sites} %
to visualize landscape hotspots (Fig.~\ref{fig:num_photos_at_site}, Tab.~\ref{tab:site_pop_table}). %
The number of unique visitors (traveler icon) and average time spent (clock icon) at each site are also shown.
Machu Picchu was unsurprisingly the most popular site ($1,498$ visitors and $31,550$ photos taken), followed by Cuzco ($1,273$ visitors and $11,842$ photos taken), as the former is one of the most iconic heritage sites globally and the latter is the major point of entry to the region. 
Of the sites visited, Pikillacta and Tipón were the least popular, likely due to their remote location to the southeast of Cuzco. 

\begin{figure*}[t]
    \begin{center}
   \includegraphics[width=.95\linewidth]{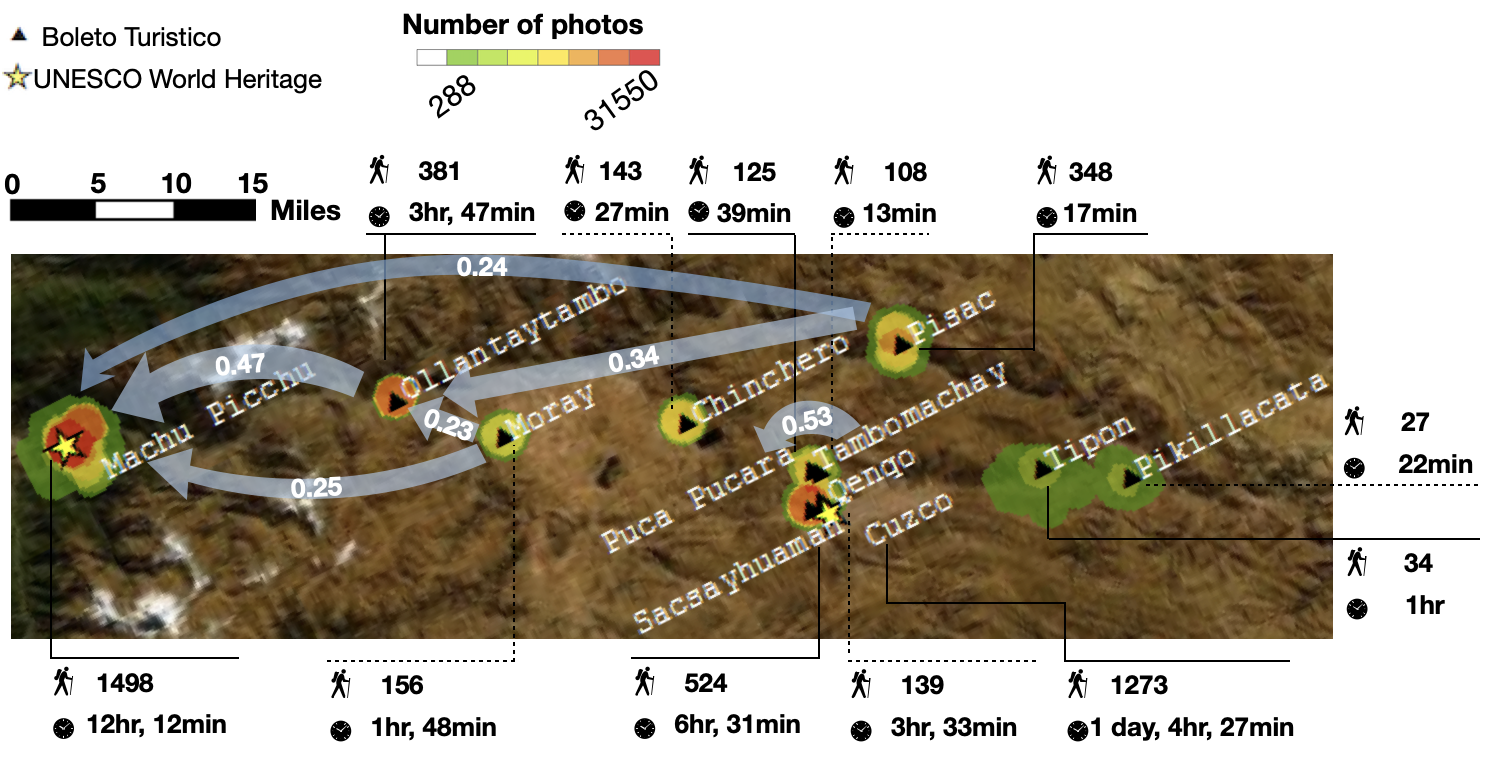}
   \vspace*{-0.05in}
    \caption{Overview of tourist movement patterns discovered in the image data. Popularity of sites measured by the unique number of tourists (shown by traveler icons) and number of photos taken per site (color coded in heatmap). The average time tourists spent per site is shown by clock icons. Top transitions between sites are on arrows labeled by probabilities.}
   \label{fig:num_photos_at_site}
   \end{center}
\end{figure*}

Popularity rankings were generally consistent between the two variables \KG{(visitor count and photo frequency)}, with slight differences between Chinchero and Moray.
Tourists spent the longest time at Cuzco, which serves as the region's only transportation hub and is where most hotel and guesthouses are located.  
Tourists were \KG{found} to spend upwards of $10$ hours at Machu Picchu.
Travel to Machu Picchu is often packaged as a day trip or overnight stay via rail, allowing for more time to be spent at the site. Tourists may also choose to trek the Inca Trail, which involves a multi-day excursion through the Andes mountains to reach Machu Picchu.
By contrast, tourists spent less than an hour at Tambomachay, Pisac, Chinchero, and Puca Pucara. %
Tour groups with limited time schedules may account for the amount of time spent at Pisac and Chinchero, whereas Tambomachay and Puca Pucara are small and can be covered quickly before moving to more popular sites like Sacsayhuaman and Qenqo.

\KG{These statistics offer a window into the cultural heritage circuit over the last 15 years in ways that are not possible with traditional methods.  For example, ticket purchase rates or manual surveys do not capture durations of visits, photographic preferences, etc., but the large-scale tourist photos do.}

\paragraph{Transition Patterns Changing over Time} %
Next we examine the movement of tourists.  We refer to our Markov matrix over all $12$ archaeological sites as the \emph{transition} matrix.
First, we compute the transition probabilities for the entire $2004$ to $2019$ time range. The top $6$ transitions with probabilities greater than $0.2$ are plotted as arrows in Fig.~\ref{fig:num_photos_at_site}.
The most frequent transition is from Tambomachay to Puca Pucara as they are only a $3$ minute walking distance. The other most frequent transitions moved toward Machu Picchu. In most cases, visitors traveled to one of four sites prior to departing for Machu Picchu: Ollantaytambo, Moray, Pisac, and Chinchero. Their proximity to Machu Picchu may account for this pattern. 

\begin{figure}[t]
    \begin{center}
   \includegraphics[width=.8\linewidth]{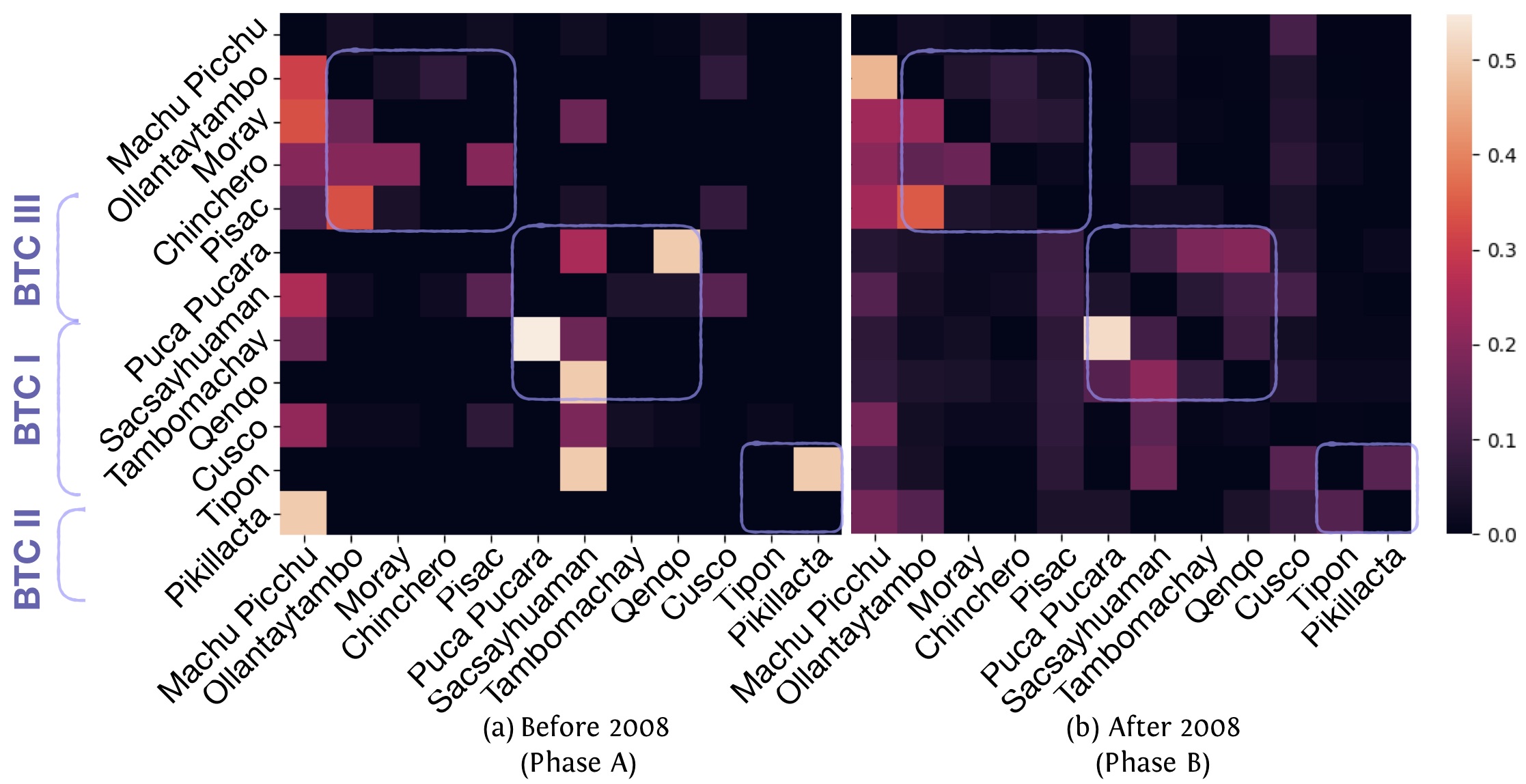}
    \caption{Transition matrix comparison of before and after the introduction of the BTC tourist package (2008). In general, transitions after 2008 become denser. Three outlined boxes represent site transition within %
    BTC I, II, III respectively. Note how transitions within each group become more dispersed after the ticket packages were released.}
   \label{fig:before_after_transition}
   \end{center}
\end{figure}

Next, we compute a separate transition matrix for travels made before (Phase A) and after $2008$ (Phase B), which marks the introduction of additional BTC tourist packages: \NP{a 1-day pass to Sacsayhuaman, Qenqo, Tambomachay, and Puca Pucara (BTC I); a 2-day pass to the archaeological sites of Tipón and Pikillacta (BTC II); and a 2-day pass to Pisac, Ollantaytambo, Chinchero, and Moray (BTC III)}. Fig.~\ref{fig:before_after_transition} shows the matrices plotted as heatmaps, where the $i^{th}$ row and $j^{th}$ column element contains the probability of moving from site $i$ to site $j$. \NP{Phase A displays higher site specific transition sequences, like Pikillacta to Machu Picchu and Puca Puca to Qenqo, and low regional diversity of movement between sites. The site specific transitions in Phase A decrease in Phase B as regional transition probabilities increase across the matrix. This suggests that there is a greater range and dispersal of site transitions in Phase B, weakening the stronger site-to-site sequences that previously existed.}  
The three outlined boxes represent BTC I, II, and III groupings, respectively. 

For the BTC I grouping, %
Phase A transitions are concentrated around nearby sites, i.e., Tambomachay to Puca Pucara or Qenqo to Sacsayhuaman. During Phase B, transition patterns become more symmetric and diffused among all four sites. This is true for transitions in BTC II and III as well, though slightly less obvious. 
The change in BTC transition patterns may indicate how the ticket packages encourage tourists to explore more sites through condensed day-trip itineraries. 
In Phase B, travel patterns shift away from BTC I sites and toward BTC III sites, probably due in part to increased hotel construction in the Sacred Valley, which has intensified the use of Ollantaytambo as the point of departure to Machu Picchu, as well as access to Chinchero, Pisac, and Moray.

Movements between sites on the BTC III ticket (Chinchero, Pisac, Ollantaytambo, Moray) were the most complex across Phase A and Phase B. A popular circuit between BTC III sites was discernible during Phase A and appeared to be the antecedent for the BTC III (2-day) pass: %
Following the addition of the BTC III (2-day) pass, visits to BTC III sites and transition patterns shift significantly.
The most significant of these impacts affected travel to Pisac. During Phase A, visitors to Pisac accessed the site from two main points: Chinchero %
or Sacsayhuaman. %
Phase B saw the disappearance of these two transition points to Pisac,
with all sites similarly likely to access it, while the main transit \emph{out} from Pisac, i.e., heading towards Ollantaytambo, remained the same in both Phases. 

\NP{Overall, the transition matrix indicates that changes to the BTC reflect existing touristic access practices as well as the altered ways that tourists might move among the sites on a given BTC ticket and how they might transition between those tickets and other heritage sites.}

\begin{figure}[t]
    \begin{center}
   \includegraphics[width=.8\linewidth]{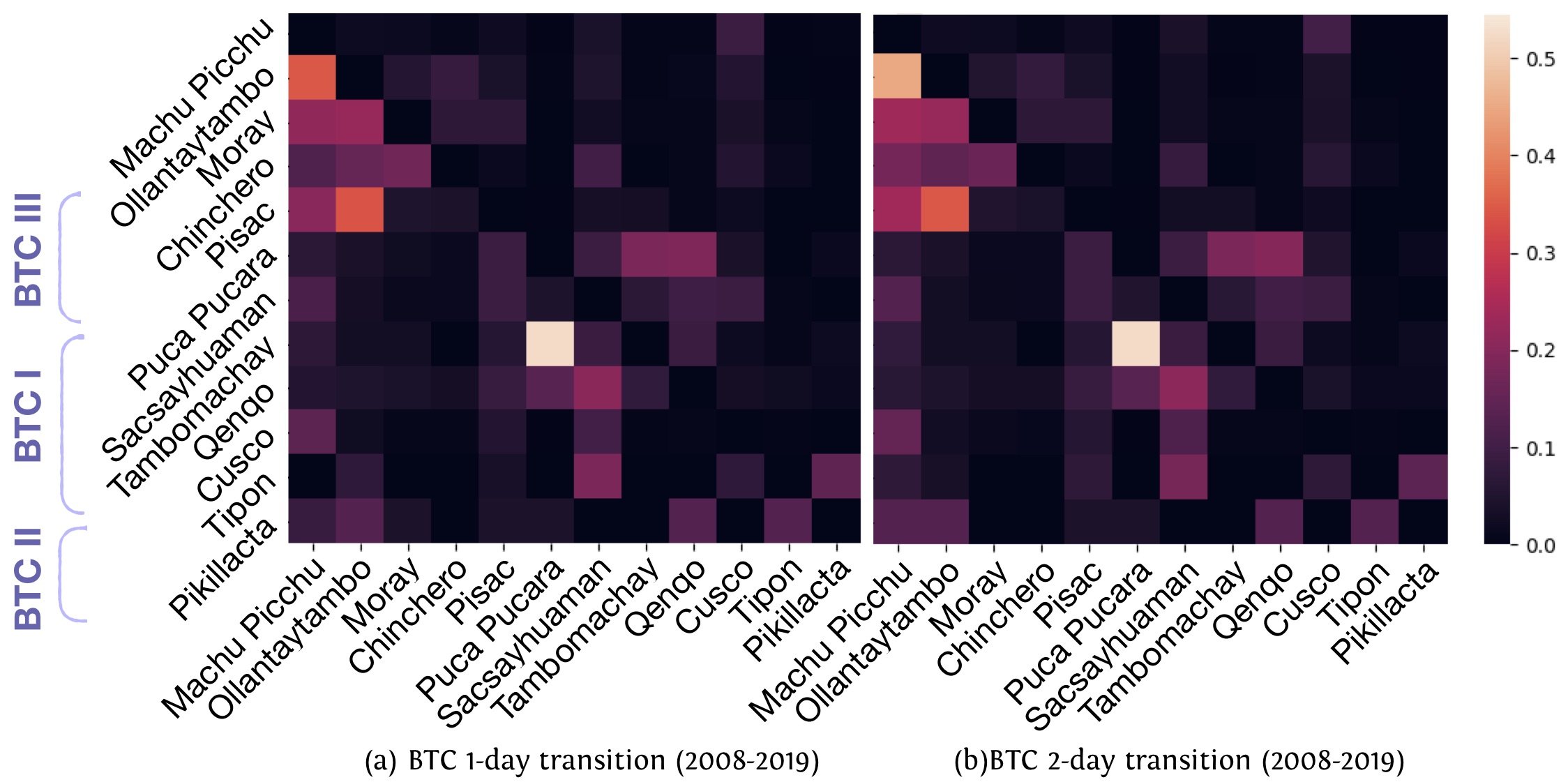}
   \vspace{-3mm}
    \caption{\KH{Comparison of transition patterns considering 1-day and 2-day passes}. \NP{Significant changes are noted in transitions from BTC III sites to Machu Picchu while observed transition probabilities of tourists traveling to Machu Picchu from BTC I sites remained low.}}
   \label{fig:1day_2day_transition}
   \end{center}
\end{figure}

\paragraph{BTC 1 and 2-day Passes}

\NP{Next we explore Phase B transition trends by isolating tourist movement within a 1- and 2-day time frame.} %
Figure \ref{fig:1day_2day_transition} represents heatmaps for 1-day and 2-day periods corresponding to the maximum time duration allowed for travel to BTC I, II, III sites during Phase B and transition probabilities greater than $10\%$. \NP{Our results show that the introduction of the 2-day pass in Phase B led to a significant shift in transitions from BTC III sites to Machu Picchu when compared to the full 10-day pass}. %
\NP{We observed a ten percent decrease in transitions from Ollantaytambo to Machu Picchu (10-day vs. 2-day) as other points of departure corresponding to BTC III (Chinchero and Pisac) saw significant increases (Tab.~\ref{tab:MachuPicchu_transition})}. \NP{This also corresponds to similar changes observed when comparing the BTC transitions in Phase A and Phase B above (Fig.~\ref{fig:before_after_transition})}. 

\begin{table*}[t] %
  \centering
  \captionsetup[subfigure]{labelformat=empty}
  \footnotesize
  \caption{\NP{Transition trends from BTC I and BTC III sites to Machu Picchu during Phase B are compared. A significant change in transition rates is seen when moving from BTC III sites to Machu Picchu (10-day pass vs. 2-day pass), while travel from BTC I sites to Machu Picchu remains relatively low regardless of 10-day vs. 1-day pass in Phase B. For the BTC I site of Sacsayhuaman, this is a departure from transition popularity in Phase A.}}
  \subfloat[Transition trends from BTC I sites to Machu Picchu] {
  \begin{tabular}{@{\extracolsep{4pt}}lc@{}}
    Site Transition      & 10 Day vs. 1 Day Pass \\
          \midrule
    Sacsayhuaman $\rightarrow$ Machu Picchu & $13\%$ vs $13\%$\\
    Qenqo $\rightarrow$ Machu Picchu & $8\%$ vs $6\%$\\
    Puca Pucara $\rightarrow$ Machu Picchu & $6\%$ vs $7\%$\\
    Tambomachay $\rightarrow$ Machu Picchu & $7\%$ vs $8\%$\\
  \end{tabular}
  }\hfill
  \subfloat[Transition trends from BTC III sites to Machu Picchu] {
  \begin{tabular}{@{\extracolsep{5pt}}lc@{}}
    Site Transition      & 10 Day vs. 2 Day Pass \\
          \midrule
    Ollaytantambo $\rightarrow$ Machu Picchu & $45\%$ vs $35\%$\\
    Pisac $\rightarrow$ Machu Picchu & $13\%$ vs $21\%$\\
    Chinchero $\rightarrow$ Machu Picchu & $5\%$ vs $12\%$\\
    Moray $\rightarrow$ Machu Picchu & $24\%$ vs $22\%$\\
  \end{tabular}
  }
  \label{tab:MachuPicchu_transition}
\end{table*}

We also observed that movement from BTC I sites to Machu Picchu compared to both the Phase B 10-day and 2-day passes is significantly lower than movement from BTC III sites (Tab.~\ref{tab:MachuPicchu_transition}).\NP{The low transition probability of BTC I sites to Machu Picchu in Phase B mainly impacted the site of Sacsayhuaman, which saw a relatively high rate of transition to Machu Picchu during Phase A and a severe decrease in transition probabilities in Phase B}. \NP{While the changes to the BTC have not affected how BTC I sites near Cuzco are accessed, the addition of the 1- and 2-day passes in Phase B created strong differences in transitions to Machu Picchu from associated BTC III sites. This is attributed in part to an increase in hotel beds in the Sacred Valley allowing tourists greater flexibility when starting and ending their BTC III days at different sites (depending on whether they are staying in Cuzco, Ollantaytambo, or Urubamba).}

\KG{Again, these findings are valuable to understanding the heritage circuit for anthropologists, and they become significantly more accessible by detecting patterns from in-the-wild photos, as we propose.}

\paragraph{Canonical View Analysis}
\KG{Next we examine how tourists are photographing these sites, and how those patterns interact with heritage-based operations.} \KH{We first detect the scene types present in each photo to obtain the distribution of captured scenes across all sites (Fig.~\ref{fig:scene_labels}). We then break this distribution down to first visualize which scenes dominate a specific site (Fig.~\ref{fig:tsne_all_sites}), and \NP{then determine} what the scenes look like at each site (Fig. \ref{fig:full_clusters}).}
\KH{Finally, we quantitatively complement our above analysis with statistics of discovered photo clusters at each site (Tab. \ref{tab:site_cluster_numbers}), and discuss our findings in detail.}

\begin{figure}[t]
   \begin{center}
   \includegraphics[width=\linewidth]{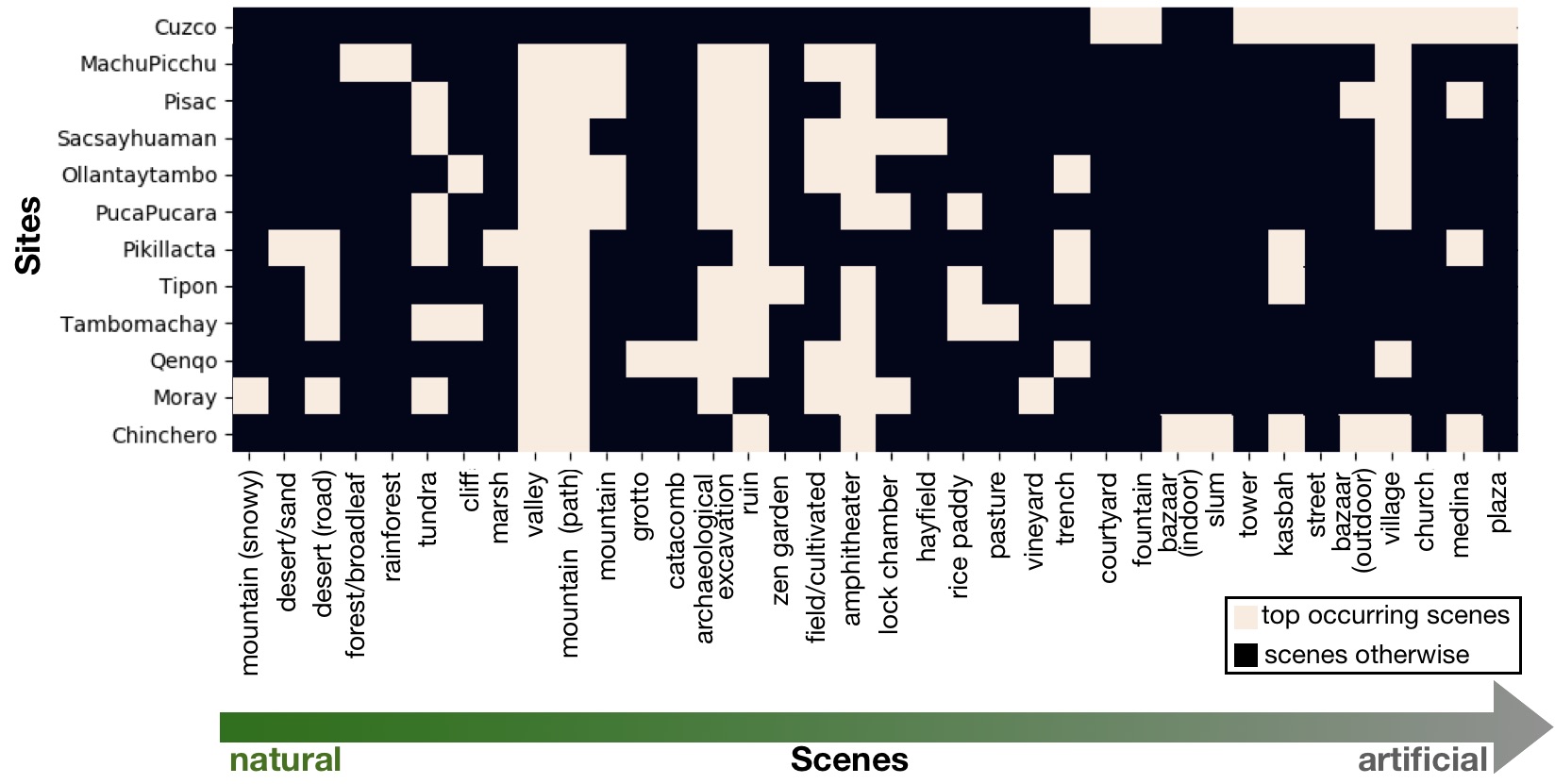}
   \caption{Scene-site occurrence matrix showing distribution of \KH{photographed} scenes across sites. The most common scene labels were: mountain path and valley (i.e. mountains), archaeological excavation and ruins (i.e. stonework), amphitheater (i.e. terracing). These statistics reveal what tourists saw and photographed most often at each of the 12 sites.} %
      \label{fig:scene_labels}
   \end{center}
\end{figure}
\NP{What do tourists notice at the sites and consider worthwhile to photograph and post online? Do tourist images converge around a set of canonical views, some of which were first introduced to popular media more than a century ago?}
\NP{To answer these questions,}
\KH{we quantify the contents of discovered canonical views per site. We use a ResNet50 \citep{resnet} pre-trained on scene categories in the MIT Places \citep{zhou2017places} dataset to predict scene labels in our dataset. Predicted scenes were aggregated on photos in each site to retrieve the most frequent scene labels. We take the most frequent $10\%$ of scenes from each site as a representative scene category, and plot this scene-site occurrence matrix (Fig.~\ref{fig:scene_labels}). The scenes are sorted from most natural (left) to most artificial (right).}
\KH{Common scenes occurring across all \NP{BTC and UNESCO sites} were: `Mountain Path' and `Valley' (i.e. mountains), `Archaeological Excavation' and `Ruins' (i.e. stonework), and `Amphitheater' (i.e. agricultural terracing).} 
\NP{Labels such as `Medina', `Kasbah', and `Bazaar' also corresponded well to market scenes captured in Cuzco, Pisac, and Chinchero.
Roadside markets at the entrance of archaeological sites are common in the Cuzco region and account for similar labels associated with Pikillacta and Tipón.} \NP{Our detected scene-labels show a strong correlation to major scene themes identified within our t-sne and cluster results below.}

\begin{figure}[t]
    \begin{center}
   \includegraphics[width=\linewidth]{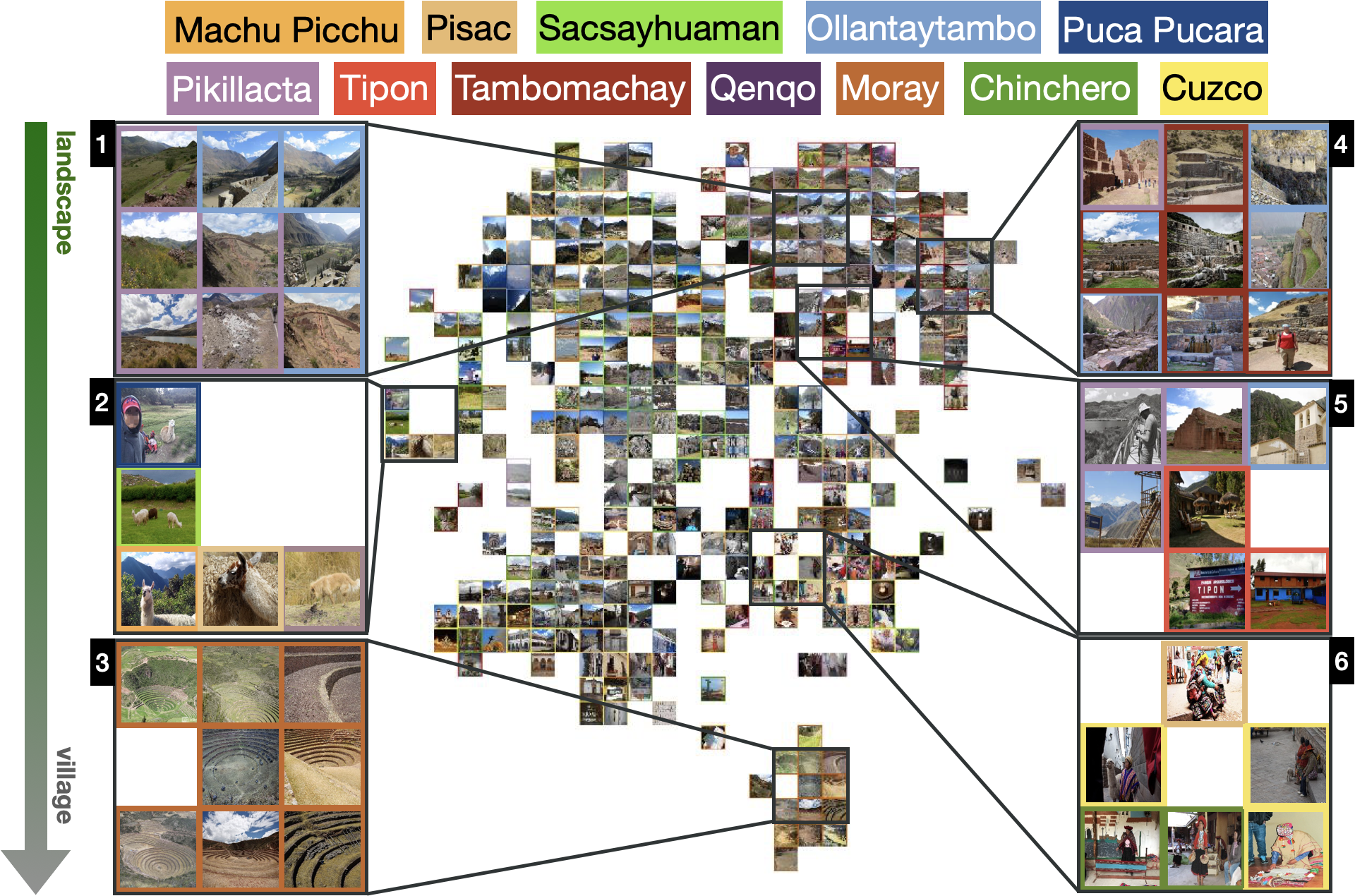}
    \caption{t-sne~\citep{maaten2008tsne} visualization of a subset of photos from all sites. Images are framed by colors corresponding to sites shown in top legend. 
    \KH{Visualization shows a gradual theme variation from landscape views (top) to village life (bottom). Bounding boxes show zoom-ins of discovered example themes (e.g. mountains, stone work, agriculture terracing). 
    Similar architectural (Box 4) and terrain aesthetics (Box 1, 3) across sites reveals shared visuality in Cuzco's heritage circuit.}
    }
   \label{fig:tsne_all_sites}
   \end{center}
\end{figure}

\KH{To visualize the photos at each \emph{scene}, we sample photos from each site, extract their CNN features \NP{using ResNet50 and ImageNet}, and run a t-sne~\citep{maaten2008tsne} visual embedding, as shown in Fig.~\ref{fig:tsne_all_sites}.} %
\NP{Vizualization of shared themes (i.e., predicted scenes) across all sites include mountain landscapes, alpaca, and stone architecture (Box 1, 2, 4) which are similar to our common scene-labels}. %
Photos of Peruvian peoples wearing traditional dress are mostly found in Cuzco, Chinchero, and Pisac (Box 6). At these locations, large open-air markets attract tourists wishing to purchase Peruvian textiles, pottery, and souvenirs. Cuzco also hosts many festivals and parades to which traditional dance groups from across the country are invited.
A high percentage of images also correspond to colonial architecture, particularly in Cuzco and Chinchero where colonial churches are a major attraction (Box 5). 

\begin{figure}[t]
    \begin{center}
   \includegraphics[width=\linewidth]{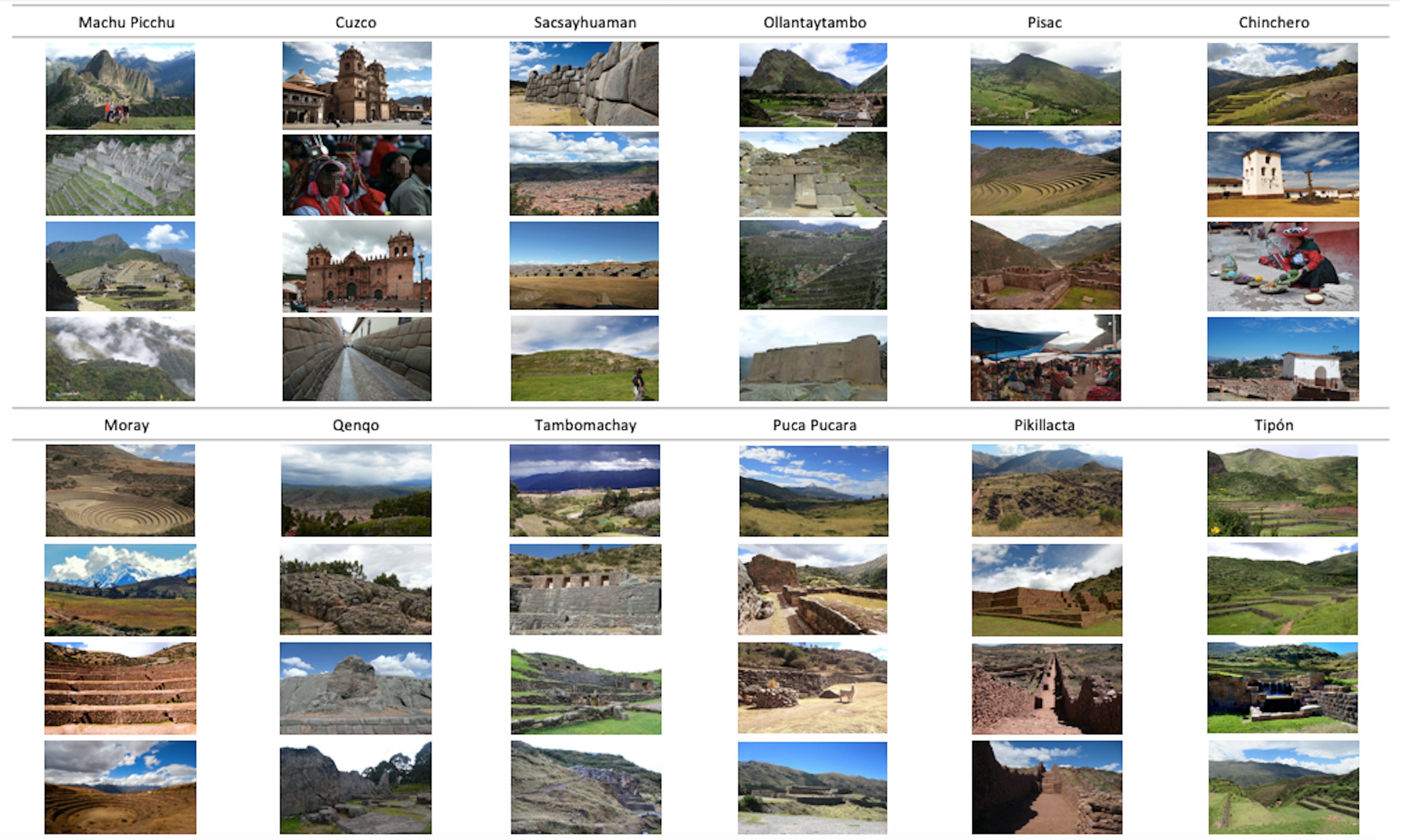}
    \caption{Representative cluster images are pictured for all 12 sites for all 57,804 photos captured by more than 2,000 unique travelers.
    \KH{Not only do popular themes (e.g. stone architecture, mountains, agricultural terracing) correspond to our distribution of detected scenes in Figure~\ref{fig:scene_labels}, but also images of site-specific scenes (e.g. colonial architecture, indigenous peoples, market scenes) are depicted \NP{for the sites of} Cuzco and Chinchero.}}
   \label{fig:full_clusters}
   \end{center}
\end{figure}

\KH{To visualize representative photos at each \emph{site}, we run affinity propagation (AP) algorithm per site to discover clusters, and take the center images from the largest clusters as representative images (Fig.~\ref{fig:full_clusters}).} Although tourist interests leaned slightly toward scene-captures associated with artificial scenes containing some manmade elements, our results showed a near-equal split between natural and artificial scene labels corresponding with our cluster themes. Scene labels also showed a strong correlation with our t-sne visualizations.
\NP{Our t-sne and cluster analysis indicates that modern tourists continue to be drawn to the same exotic viewsheds depicted in early expeditionary images: Andean landscapes (e.g. terracing), Inca ruins, indigenous people, and Colonial architecture (Fig.~\ref{fig:historic_modern_view}).}

To quantitatively understand the silhouette of clusters discovered at each site, we look at three measurements: number of clusters discovered, separation score between clusters, and the compactness score within a cluster. The \KG{first two} measure the \emph{diversity} of clusters, as the more clusters discovered and the more separated clusters are, the more different clusters are from each other---\KG{and hence the more variety is present in how tourists photographed the given site}. The compactness score measures the \emph{homogeneity} of clusters, as the tighter or denser a cluster is, the more similar data points are to each other within that cluster---\KG{and hence the less variation in tourist photos of a particular landmark}.
The separation score is computed as the visual distances between cluster exemplars, averaged over all pairwise clusters. The compactness score is computed as the maximal pairwise visual distance between data points in the cluster, averaged over all clusters.\footnote{\KG{As} AP clustering %
provides few %
large clusters and many small clusters, we 
\KG{compute separation and compactness using} %
the top $10\%$ largest clusters.}

\begin{table*}[t] %
  \centering
  \captionsetup[subfigure]{labelformat=empty}
  \footnotesize
  \caption{Clustering results on all sites. $\#$ clusters show the number of clusters discovered at each site by AP clustering.  (Inter-cluster) separation shows how well separated a cluster is from other clusters. These together measure how different the clusters are; higher means more different. (Intra-cluster) compactness shows the average homogeneity of a cluster; lower means more homogeneous. \KH{UNESCO sites have the greatest amount of clusters \NP{(meaning more varied photos taken) and reflect how tourists spent longer periods of time at these sites}. Large sites (Machu Picchu and Ollantaytambo) have the highest compactness scores, revealing the reproduction of popular photo themes, while smaller sites (Chinchero, Puca Pucara, and Qenqo) have lower compactness, likely due to unrestricted tourist movement there. See text for detailed discussion.}}
  \subfloat[] {
  \begin{tabular}{@{\extracolsep{4pt}}c|p{8mm}ccccc@{}}
    site          & \parbox{8mm}{Machu\\Picchu} & Cuzco   & Chinchero        & Moray         & Ollantaytambo & Pisac\\
          \midrule
    $\#$ clusters $\uparrow$ & \textbf{1425}     &  \textbf{642}     & 75      & 67    & \textbf{279}          & 216 \\
    separation $\uparrow$  & 23.24     & \textbf{25.03}   & \textbf{25.08}    & \textbf{25.39} & 21.96       & $22.33$  \\
    compactness $\downarrow$   & \textbf{20.67}  &  23.47  & 26.03           & $21.80$        & \textbf{20.46} & $22.42$      \\
  \end{tabular}
  }\hfill
  \subfloat[] {
  \begin{tabular}{@{\extracolsep{4pt}}c|cp{8mm}cccc@{}}
    site   & \parbox{8mm}{Puca\\Pucara} & Qenqo  & Sacsayhuaman & Tambomachay & Pikillacta & Tipón \\
          \midrule
    $\#$ clusters $\uparrow$  & 48    & 59    &          257          & 47         & 36        & 56 \\
    separation $\uparrow$     &       $21.26$    &$23.05$ &  $23.68$     & $20.92$     & $21.93$    & 22.39 \\
    compactness $\downarrow$  &         $24.06$         &$24.31$ & $21.19$    &    \textbf{19.84}    & $22.07$    & $21.37$    \\
  \end{tabular}
  }
  \label{tab:site_cluster_numbers}
\end{table*}

Table \ref{tab:site_cluster_numbers} presents the  quantitative \KG{metrics}.
Cluster sizes range from $364$ to $1,425$,
with $114$ being the average for BTC sites and $1,034$ for UNESCO World Heritage sites. 
Overall, UNESCO sites have more diverse clusters, as measured by the number of clusters and separation score, reflecting the larger transition time at these sites, whereby people are spending more time, and taking more \KG{varied} images.
Conversely, Machu Picchu and Ollantaytambo have the highest compactness scores,
revealing that the reproduction of popular photo themes and viewpoints are more likely at large sites. %

\KG{These patterns shed light on the impact of cultural heritage management decisions in recent years.} 
Due to the growth of mass tourism, several of Cuzco's most popular archaeological sites are no longer `free-range'. Instead, movement \emph{within} these sites is restricted to a predefined path. For example, at Machu Picchu visitors must purchase their site permits months in advance to guarantee entry before they travel. Upon arrival to the site, tourists are guided through a set route to avoid damage to sensitive areas. %
\KG{In fact,} the governing body of UNESCO \KG{continues to express} concerns over the volume of tourists accessing the site. In $2019$, new regulations were adopted limiting tourists to a maximum of $4$ hours at the site, with three daily entry shifts scheduled between $6$ am and $3$ pm. %
Efforts to prevent the deterioration of heritage spaces and restrict movements have also been initiated at Ollantaytambo where
tourists are directed up the Terraces of Pumatallis toward the Sun Temple and Wall of Six Monoliths before moving toward the site's funerary complexes. %
\KG{Our results suggest that} these conservation initiatives and entry regulations have increased photo homogeneity at larger sites as tourists are made to follow predefined routes through heritage attractions, allowing for the reproduction of objects and scenes.  %
At smaller sites like Chinchero, Puca Pucara, and Qenqo, photo homogeneity remained lower as tourist movement is unrestricted, allowing a wider variety of images.

\begin{figure}[t]
    \begin{center}
   \includegraphics[width=\linewidth]{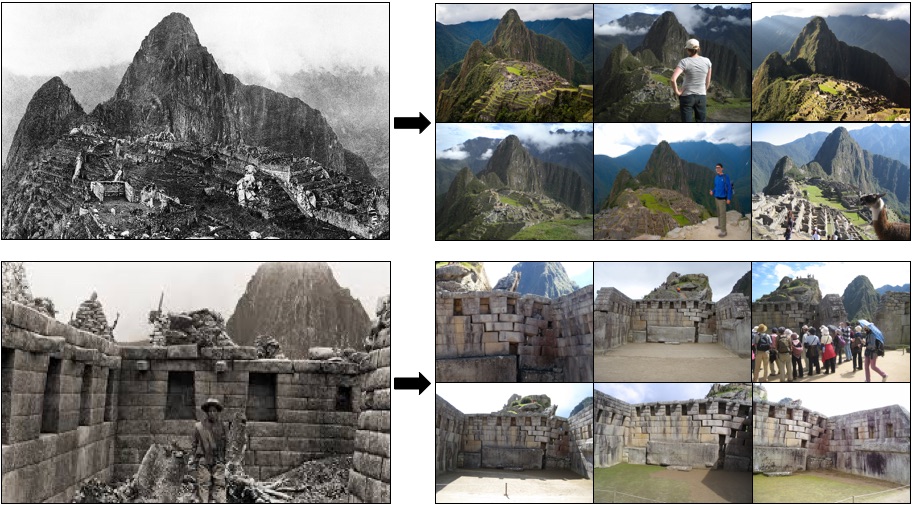}
   \caption{Left: Historic photographs taken by Hiram Bingham during the Yale Peruvian Expedition (1911-1915). Right: \NP{User images from our internet photo dataset} within the last 15 years (2004-2019). \NP{Images taken by Bingham show a strong visual influence on preferred scene-captures and objects photographed by tourists.}}
      \label{fig:historic_modern_view}
   \end{center}
   \vspace*{-0.25in}
\end{figure}
\KG{Finally, our results hint at the influence of historical images in guiding how current tourists experience a heritage site.  Notably,} many of the representative images identified in the top $10\%$ of clusters share aesthetic qualities with historic photographs published by Hiram Bingham in the early 20th Century (Fig.~\ref{fig:historic_modern_view}).
The continued reproduction and distribution of these scenes on social media platforms creates a visual heritage narrative that influences the expectations of future travelers on the landscape. When a particular image or scene achieves dominance, alternative ways of experiencing landscapes are inevitably obscured \citep{waterton2009sights}. Therefore, heritage discourses, like that generated by early explorers and perpetuated by Cuzco's modern tourism industry, generate aesthetic legacies over time by reproducing that which is meant to be seen based on perceived experiences and values extracted from audience consumption (aesthetic, economic, cultural, etc. \citep{watson2010waterton}).

\section{Conclusion}
\vspace*{-0.05in}
Our study provides an innovative (and first of its kind) application of computer vision and machine learning algorithms to quantify the visuality of heritage landscapes and analyze the influence of heritage regulations on tourist circuits in the Cuzco region. By utilizing publicly available, geotagged source data from \NP{internet photos} we are able to analyze tourist movement amongst BTC and UNESCO sites. Knowledge of travel patterns across heritage landscapes provides vital information not only for heritage conservation and the management of archaeological sites, but also for assessing the economic impacts that new regulations may have on local communities who depend on the recurrent influx of visitors. 

Through \KG{pattern recognition tools} like feature extraction, \KG{image recognition}, clustering, and t-sne \KG{embeddings}, we are able to show a broad overview of the diversity and homogeneity of tourist generated images and identify how travelers are visually experiencing Cuzco's heritage landscapes. These techniques also provide opportunities for qualitative and quantitative comparisons with historic imagery to address the reproduction of aesthetic legacies. The power of convolutional neural networks (CNNs) to extract image features and compare tens of thousands of photos within a short period, as well as the ability of affinity propogation (AP) to automatically cluster site images based on their visual characteristics \KG{discovers trends at a scale that would be} extremely difficult and time consuming, \KG{if not impossible,} through human labor alone.

\NP{The methods and techniques that we have adapted for our regional study of Cuzco's heritage circuit can be generalized and scaled globally to study different heritage sites and time periods. The interdisciplinary nature of our approach and the unique insights we are able to generate with machine learning are significant for developing sustainable heritage practices as mass tourism increasingly threatens heritage sites worldwide.} Future heritage research that could benefit from our %
\KG{ideas} include: examining how tourist scene-captures have evolved over time, studying movement patterns and hotspots \emph{within} heritage sites, and identifying outlier archaeological sites `on-the-rise' through social media platforms %
and their potential economic influences.

\vspace{7mm}
\paragraph{Acknowledgements}
UT Austin is supported in part by NSF IIS-1514118.

\end{frontmatter}

\bibliographystyle{apalike}
\bibliography{arc,cvml}

\end{document}